\documentclass[journal]{IEEEtran}

\ifCLASSINFOpdf
\else
   \usepackage[dvips]{graphicx}
\fi
\usepackage{url}

\usepackage{graphicx}
\usepackage{settings}

\begin{document}

\title{MOV-Modified-FxLMS algorithm with Variable Penalty Factor in a Practical Power Output Constrained Active Control System}

\author{
    Chung~Kwan~Lai, 
    Dongyuan~Shi, \IEEEmembership{Member, IEEE}, 
    Bhan~Lam, \IEEEmembership{Member, IEEE}, 
    Woon-Seng~Gan, \IEEEmembership{Senior Member, IEEE}

    
    \thanks{All authors are with the  
    School of Electrical \& Electronic Engineering, Nanyang Technological University (NTU), Singapore. Email: \{chungkwan.lai, dongyuan.shi, bhanlam, \mbox{ewsgan}\}@ntu.edu.sg.}
}

\markboth{Submitted to Signal Processing Letters}
{Lai \MakeLowercase{\textit{et al.}}: MOV-Modified-FxLMS algorithm with Variable Penalty Factor in a Practical Power Output Constrained Active Control System}
\maketitle
\begin{abstract}
Practical Active Noise Control (ANC) systems typically require a restriction in their maximum output power, to prevent overdriving the loudspeaker and causing system instability. Recently, the minimum output variance filtered-reference least mean square (MOV-FxLMS) algorithm was shown to have optimal control under output constraint with an analytically formulated penalty factor, but it needs offline knowledge of disturbance power and secondary path gain. The constant penalty factor in MOV-FxLMS is also susceptible to variations in disturbance power that could cause output power constraint violations. 
This paper presents a new variable penalty factor that utilizes the estimated disturbance in the established Modified-FxLMS (MFxLMS) algorithm, resulting in a computationally efficient MOV-MFxLMS algorithm that can adapt to changes in disturbance levels in real-time. Numerical simulation with real noise and plant response showed that the variable penalty factor always manages to meet its maximum power output constraint despite sudden changes in disturbance power, whereas the fixed penalty factor has suffered from a constraint mismatch.

\end{abstract}


\IEEEpeerreviewmaketitle

\section{Introduction} \label{sec:Introduction}

\IEEEPARstart{A}{ctive} Noise Control (ANC) often utilizes adaptive algorithms to reduce acoustic pressure at a desired location \citep{Elliott2000a, Kuo1996, Kajikawa2012, Chang2016, Qiu2019}. The widely used filtered-reference least mean square (FxLMS) algorithm uses stochastic steepest gradient descent to iteratively update control filter coefficients and minimize the sum of squared error signals over time. When suppressing loud disturbances, however, the control signal amplitude may exceed the rated system output power, resulting in output saturation that introduces non-linearities \citep{Jardin2007, Heinle1998}. Thus, a practical ANC system should regulate its output within the rated power to prevent overdriven loudspeakers, which introduce distortion and may lead to mechanical failure.
Moreover, signal clipping at the audio power amplifier can cause distortion, divergence in the FxLMS algorithm due to overflow in the controller coefficients and its non-linear behavior \citep{Shi2017, Qiu2001}, which should be considered in realistic adaptive models \citep{ Sahib2012, Ghasemi2016, Russo2007, Tobias2002}, though often ignored for simplicity \citep{Costa2001}. Clipping at the microphone pre-amplifier will limit the algorithm's convergence rate, but will have minimal impact on control performance \citep{Kuo2004}.

Various approaches exist to limit the output signal during adaptive control \citep{Elliott1996, Ahmed2021, Qiu2001, HuiLan2002, Shi2018, Shi2019a, Shi2021b, Taringoo2006, Shi2022}. The rescaling \citep{Qiu2001, HuiLan2002} and two-gradient direction FxLMS \citep{Shi2018} algorithms adjust the control filter coefficients when the output exceeds a certain threshold. While effective, constraining the output signal amplitude is less practical than constraining its output power as the peak amplitude varies with different noise types. It is thus difficult to pre-determine an amplitude threshold, which we usually set as the rated power, whereas a reliable power constraint threshold can be obtained from an audio equipment datasheet (e.g. amplifier). Therefore, employing a power constraint is a more practical approach to ensure a safe and reliable operation of an amplifier. The optimal leaky FxLMS (OLFxLMS) algorithm \citep{Shi2019a, Shi2021b} uses an optimal regularisation parameter to effectively constrain the output power, but this comes with a trade-off of high computational cost \citep{Shi2021b}. 

In contrast, the minimum output variance FxLMS (MOV-FxLMS) algorithm \citep{Taringoo2006} limits the output variance by including a penalty factor in the cost function at a lower computational complexity. The analytical formulation of the optimal penalty factor for optimal control, under a given output variance constraint, can be derived through a quadratically constrained quadratic program problem (QCQP) and estimated based on the adaptive inverse modeling technique \citep{Shi2022}. This optimal value requires prior knowledge of the noise type, in particular, the variance of the disturbance signal, which can only be measured offline before the control stage. Additionally, it assumes constant primary noise variance during measurement and control stages, making it sensitive to any mismatch in primary noise variance, which will be illustrated in \sref{sec:SimulationResult}. 

These limitations can be mitigated if the disturbance signal is fed back to the control system, which allows insight into the disturbance variance during the control stage. Updated knowledge of the disturbance could be leveraged from the widely known modified FxLMS algorithm (MFxLMS), where the disturbance is re-estimated by removing the estimated anti-noise signal from the `error' sensors in every iteration \citep{Elliott2000a, Rupp1996}. This letter presents the minimum output variance MFxLMS (MOV-MFxLMS) algorithm in \sref{sec:Modiied_MOV_adaptive_algorithm}, along with the optimal and time-varying penalty factor for MOV-MFxLMS in \sref{sec:QPCP} and \sref{sec:EstimationPenalty}, respectively, which allows the optimal penalty factor to be obtained in real-time. This method is verified through numerical simulation in \sref{sec:SimulationResult}, before concluding its overall effectiveness in \sref{sec:Conclusion}.
\section{Minimum Output Variance Modified-FxLMS Algorithm} \label{sec:Modiied_MOV_adaptive_algorithm}
\begin{figure}[!t] \centering 
    \includegraphics[width=0.9\linewidth]{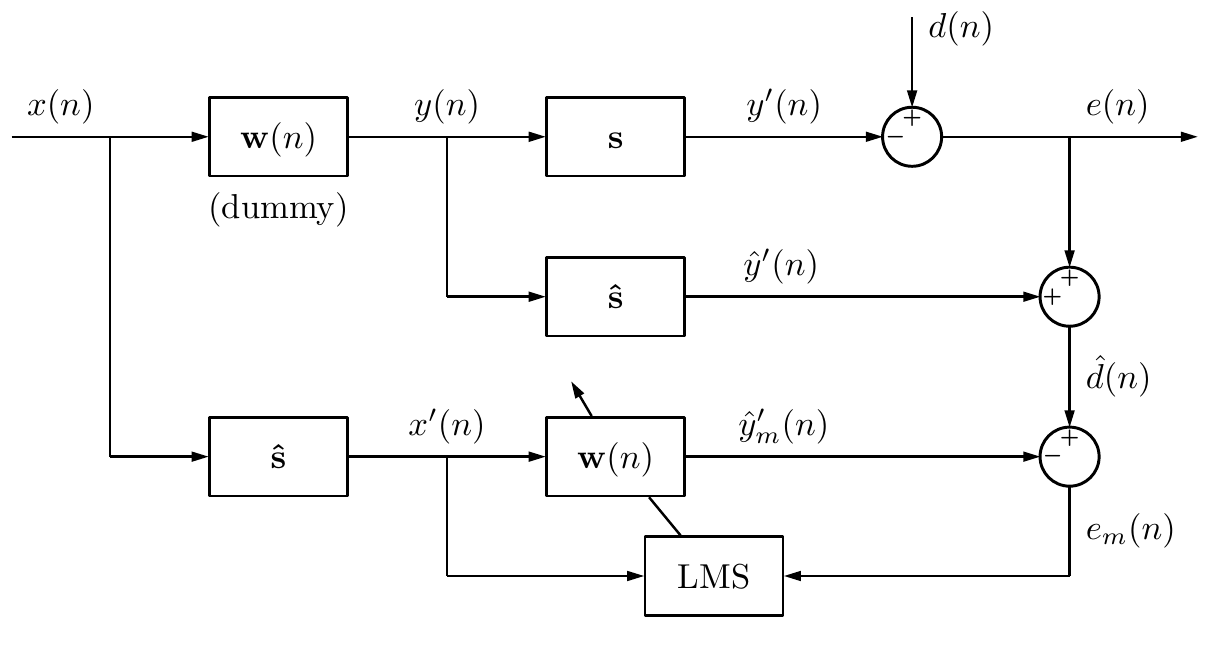} \vspace{-1\baselineskip}
    \caption{Modified-FxLMS (MFxLMS) block diagram. \label{fig:MFxLMS_blockDiagram}} \vspace{-0.5\baselineskip}
\end{figure}

\Fref{fig:MFxLMS_blockDiagram} shows a single channel adaptive feedforward ANC system using a modified-FxLMS arrangement \citep{Rupp1996, Elliott2000a}. The control filter $\mathbf{w}(n) = \left[w_0(n) \: w_1(n) \: \cdots \: w_{I - 1}(n)\right]^\mathrm{T}$ containing $I$ FIR coefficients processes the reference signal $x(n)$ to generate a control signal $y(n)$, which then propagates through the secondary path $\mathbf{s}$ to reach the error sensor forming an anti-noise signal $y^\prime(n)$. The error signal $e(n)$ is then the difference between $y^\prime(n)$ and the disturbance $d(n)$. The cost function is therefore formulated based on the minimum output variance to constrain the power output as in \citep{Elliott2000a}
\begin{align}
    J = \mathbb{E}\left[e_m^2(n) \right] + \alpha \mathbb{E}\left[ y^2(n) \right], \label{eqn:J}
\end{align}
where $\mathbb{E}[\cdot]$ and $\alpha \; (\alpha\geq 0)$ represent the expectation operator and penalty factor, respectively. The modified error signal $e_m(n)$, where subscript `$m$' is short for `modified', is expressed as
\begin{align}
    e_m(n) &= \hat{d}(n) - \hat{y}^\prime_m(n),   \label{eqn:e_m}
\end{align}
where the estimated disturbance signal $\hat{d}(n)$ and modified anti-noise signal $\hat{y}^\prime_m(n)$ are given respectively by
\begin{align}
    \hat{d}(n) &= e(n) + \sum_{l = 0}^{L - 1} \hat{s}_{l} \mathbf{w}^\mathrm{T}(n-l)\mathbf{x}(n-l) \label{eqn:d_hat_n}
\end{align}
and
\begin{align}
    \hat{y}^\prime_m(n) = \sum_{i = 0}^{I - 1}  \sum_{l = 0}^{L - 1} w_{i}(n) \hat{s}_{l} x(n-i-l),
\end{align}
while $\mathbf{{s}} = [{s}_0 \: {s}_1 \: \cdots \: {s}_{L-1}]^\mathrm{T}$, $\mathbf{\hat{s}} = [\hat{s}_0 \: \hat{s}_1 \: \cdots \: \hat{s}_{L-1}]^\mathrm{T}$ denote the secondary path and its estimation containing $L$ FIR coefficients. The control signal $y(n)$ and the filtered reference signal $x^\prime(n)$ are given by
\begin{equation}
    \begin{cases}
     y(n) &= \mathbf{w}^\mathrm{T}(n)\mathbf{x}(n),\\
     x^\prime(n) &= \sum_{l = 0}^{L - 1} \hat{s}_{l}x(n-l).
    \end{cases}\label{eqn:rhat_n}
\end{equation}
By setting the derivative of the cost function in \eref{eqn:J} to 0, the optimal control filter is derived as
\begin{align}
    \mathbf{w}_\mathrm{opt} = (\mathbf{R}_{x^\prime x^\prime} + \alpha \mathbf{R}_{xx})^{-1}\mathbf{r}_{x^\prime \hat{d}}, \label{eqn:optimalFilter_MoV}
\end{align}
where the auto-correlation matrices and correlation vector are given by 
\begin{equation} 
    \begin{cases}
        \mathbf{R}_{xx}&= \mathbb{E}[\mathbf{x}(n)\mathbf{x}^\mathrm{T}(n)],\\
        \mathbf{R}_{x^\prime x^\prime} &= \mathbb{E}[\mathbf{x}^\prime(n)\mathbf{x^\prime}^\mathrm{T}(n)],\\
        \mathbf{r}_{x^\prime \hat{d}} &= \mathbb{E}[\mathbf{x^\prime}(n)\hat{d}(n)].
    \end{cases}
\end{equation}
According to the stochastic gradient descent method to minimize the instantaneous value of \eref{eqn:J}, we can use the recursive formula for the control filter as 
\begin{align}   
    \mathbf{w}(n+1) = \mathbf{w}(n) + \mu \left[ \mathbf{x^\prime}(n)e_m(n) - \alpha \mathbf{x}(n)y(n) \right], 
    \label{eqn:MoV-MFxLMS-1}
\end{align}
which is henceforth known as the minimum output variance modified-FxLMS (MOV-MFxLMS) algorithm, with $\mu$ as the step-size. Choosing the right value of penalty factor $\alpha$ from \eref{eqn:optimalFilter_MoV} is crucial as it affects the attenuation performance. 
A large $\alpha$ will overemphasize the minimization of the $\mathbb{E}\left[y^2(n)\right]$ term in \eref{eqn:J}, resulting in a gain reduction in the control filter and a decrease in attenuation performance. On the other hand, a small $\alpha$ will not effectively constrain the power output, leading to output saturation.
\section{Optimal penalty factor for MOV-MFxLMS} \label{sec:QPCP}
The optimal penalty factor that satisfies a power output constraint can be derived by relating the optimal control filter from \eref{eqn:optimalFilter_MoV} and the optimal control filter derived from the Lagrangian optimization problem \citep{Shi2022}. The optimization problem under power output constraint is formulated as 
\begin{subequations}
\begin{alignat}{3}
    &\min_{\mathbf{w}} \; \; && \xi(\mathbf{w}) &&= \mathbb{E}\left[e_m^2(n)\right] \label{eqn:opt_a} \\
    &\;s.t.  && g(\mathbf{w}) &&= \mathbb{E}\left[y^2(n)\right] - \rho^2 \leq 0 \label{eqn:opt_b}
\end{alignat}
\end{subequations}
where $\rho^2$ is the maximum control signal output power allowed. The Lagrangian function \citep{boydConvexOptimization2004} can thus be formulated to give
\begin{align}
    \mathcal{L}(\mathbf{w}, \lambda) =  \mathbb{E}\left[e_m^2(n)\right] + \lambda \left\lbrace \mathbb{E}\left[y^2(n)\right] - \rho^2 \right\rbrace,
\end{align}
where $\lambda\;(\lambda \geq 0)$ denotes the Lagrange multiplier. By setting the derivative of the Lagrangian function to be zero, i.e. $\nabla \mathcal{L}(\mathbf{w}, \lambda) = 0$, the optimal control filter can be derived as
\begin{align}
    \mathbf{w}_\mathrm{o} = (\mathbf{R}_{x^\prime x^\prime} + \lambda_\mathrm{o} \mathbf{R}_{xx})^{-1}\mathbf{r}_{x^\prime \hat{d}}, \label{eqn:optimalFilter_Lagrange}
\end{align}
where the optimal Lagrange factor $\lambda_\mathrm{o}$ yields solutions respective to the constrained and unconstrained conditions, that is
\begin{align}
    \lambda_\mathrm{o} = \frac{\mathbb{E}\left[\hat{d}(n) \hat{y}^\prime_{m\mathrm{o}}(n)\right] - \mathbb{E}\left[\hat{y}^{\prime 2}_{m\mathrm{o}}(n)\right]}{\rho^2} \;\text{or}\; 0, \label{eqn:lambda_opt}
\end{align}
with $\hat{d}(n)$ and $\hat{y}^{\prime 2}_{m\mathrm{o}}(n)$ from \eref{eqn:lambda_opt} replace the $d(n)$ and $y^{\prime 2}_{\mathrm{o}}(n)$ terms from the MOV-FxLMS formulation, respectively. Notably, the optimal control filter in \eref{eqn:optimalFilter_MoV} is comparable to the Lagrangian formulation in \eref{eqn:optimalFilter_Lagrange} when $\mathbf{w}_\mathrm{opt} = \mathbf{w}_\mathrm{o}$, which in turn relates to the optimal penalty factor $\alpha_\mathrm{o} = \lambda_\mathrm{o}$. To remove the cross-correlation $\mathbb{E}[\hat{d}(n) \hat{y}^\prime_{m\mathrm{o}}(n)]$ from \eref{eqn:lambda_opt}, we impose an additional condition suitable in an ANC system where $\hat{d}(n)$ and $\hat{y}^\prime_{m\mathrm{o}}(n)$ are positively correlated \citep{Shi2021b, Shi2022}:
\begin{align}
    \frac{\mathbb{E} \left[\hat{d}(n)\hat{y}^\prime_{m\mathrm{o}}(n) \right]}{\sqrt{\mathbb{E}[\hat{d}^2(n)]\mathbb{E}[\hat{y}^{\prime 2}_{m\mathrm{o}}(n)]}} = 1 ,  \label{eqn:correlation}
\end{align}
where $\mathbb{E}[\hat{d}(n)] = \mathbb{E}[\hat{y}^{\prime}_{m\mathrm{o}}(n)] = 0$. The optimal penalty factor is then derived as
\begin{align}
    \alpha_\mathrm{o} = \lambda_\mathrm{o} =  G_s\left(\sqrt{\frac{\sigma_{\hat{d}}^2}{\rho^2 G_s}}-1\right), \label{eqn:optimal_penalty_factor}
\end{align}
where $G_s$ signifies the power gain of the secondary path \citep{Shi2022} given by 
{\small
\begin{align}
    G_s = \frac{\sigma^2_{\hat{y}^{\prime}_{m\mathrm{o}}}}{\rho^2} \label{eqn:Gs},
\end{align}
}\noindent
and $\sigma_{\hat{d}}^2 = \mathbb{\mathbb{E}}[\hat{d}^2(n)]$, $\sigma^2_{\hat{y}^{\prime}_{m\mathrm{o}}} = \mathbb{E}[ \hat{y}^{\prime 2}_{m\mathrm{o}}(n)]$ denote the variance of the estimated disturbance and optimally constrained anti-noise, respectively. 

As $\hat{d}(n)$ is known in the modified arrangement, the optimal penalty factor \eref{eqn:optimal_penalty_factor} can be predicted over time and therefore enabling the controller to attain the output power constraint, even if the noise type or amplitude changes throughout the control process. A variable penalty factor should theoretically outperform the pre-estimated factor in the MOV-FxLMS algorithm, which is sensitive to the variation in the disturbance.

\section{Time-varying penalty factor for MOV-MFxLMS} \label{sec:EstimationPenalty}
To forecast the ideal variable penalty factor, we need to retreive the varying $G_s$ parameter. However, it is impossible to calculate $G_s$ through \eref{eqn:Gs} as the constrained optimal anti-noise signal $\hat{y}^{\prime}_{m\mathrm{o}}(n)$ is unobtainable in practice. Additionally, the value of $G_s$ depends on the type of noise. One solution is to estimate $G_s$ through the inverse modelling technique \citep{Shi2021b, Widrow1985}, but this method is not suited for real-time implementation due to its intensive computations and slow convergence in finding the inverse path.

Alternatively, we can estimate $G_s$ using the filtered reference signal and the reference signal, as the reference signal contains the same frequency components as the disturbance in most real-world ANC applications. A more practical estimate of the secondary path's power gain is thus given by
\begin{align}   
    \hat{G}_s(n) =\frac{\sigma_{x^\prime}^2(n)}{\sigma_{x}^2(n)} \approx G_s. \label{eqn:Gs_new}
\end{align}
With the inclusion of $x(n)$ and $x^\prime(n)$, $\hat{G}_s(n)$ updates accordingly to changes in the disturbance. Furthermore, as $x^\prime(n)$ and $x(n)$ must be obtained to perform the MFxLMS algorithm, no additional convolution is required to obtain $\hat{G}_s(n)$ as opposed to the previously proposed inverse modelling technique. By assuming local stationarity in $x(n)$ and $\hat{d}(n)$ within $K$ samples, $G_s$ can be estimated in a real-time implementation as
\begin{align}
    \hat{G}_s(n) = \frac{\max \left\lbrace\sum_{k=0}^{K-1}x^{\prime 2}(n-k), \varepsilon_1 \right\rbrace}{\max \left\lbrace\sum_{k=0}^{K-1}x^2(n-k), \varepsilon_2 \right\rbrace}, \label{eqn:Gs_est_new}
\end{align}
followed by the variable penalty factor given by
{\small
\begin{align}
    \alpha (n) = \max\left\lbrace \hat{G}_s(n) \left( \sqrt{\frac{\sum_{k=0}^{K-1}\hat{d}^{2}(n-k)}{K \rho^2 \hat{G}_s(n)}} - 1\right), 0 \right\rbrace , \label{eqn:OptimalPenaltyFactorVary} 
\end{align}
}\noindent
where $\varepsilon_1$ and $\varepsilon_2$ are the minimal threshold parameters ensuring that numerator and denominator are non-zero. To prevent $\alpha(n)$ from becoming negative during the control stage, a minimum threshold of 0 will be required in \eref{eqn:OptimalPenaltyFactorVary}. Thus, substituting \eref{eqn:OptimalPenaltyFactorVary} into \eref{eqn:MoV-MFxLMS-1} will formulate the MOV-MFxLMS algorithm with a time-varying optimal penalty factor to allow robustness against variations in disturbance noise power, which is
\begin{align}   
    \mathbf{w}(n+1) = \mathbf{w}(n) + \mu \left[ \mathbf{x^\prime}(n)e_m(n) - \alpha (n) \mathbf{x}(n)y(n) \right]. \label{eqn:MoV-MFxLMS-2}
\end{align}

\section{Numerical Simulation} \label{sec:SimulationResult}
In this section, we conducted a numerical simulation comparison between the proposed variable penalty factor MOV-MFxLMS and other algorithms under different conditions. 
\subsection{Convergence path due to a sudden change in disturbance} \label{sec:SimulationResult2Tap}
\begin{figure}[!t] \centering
    \begin{subfigure}{0.49\linewidth}    \centering
    \includegraphics[width=\linewidth]{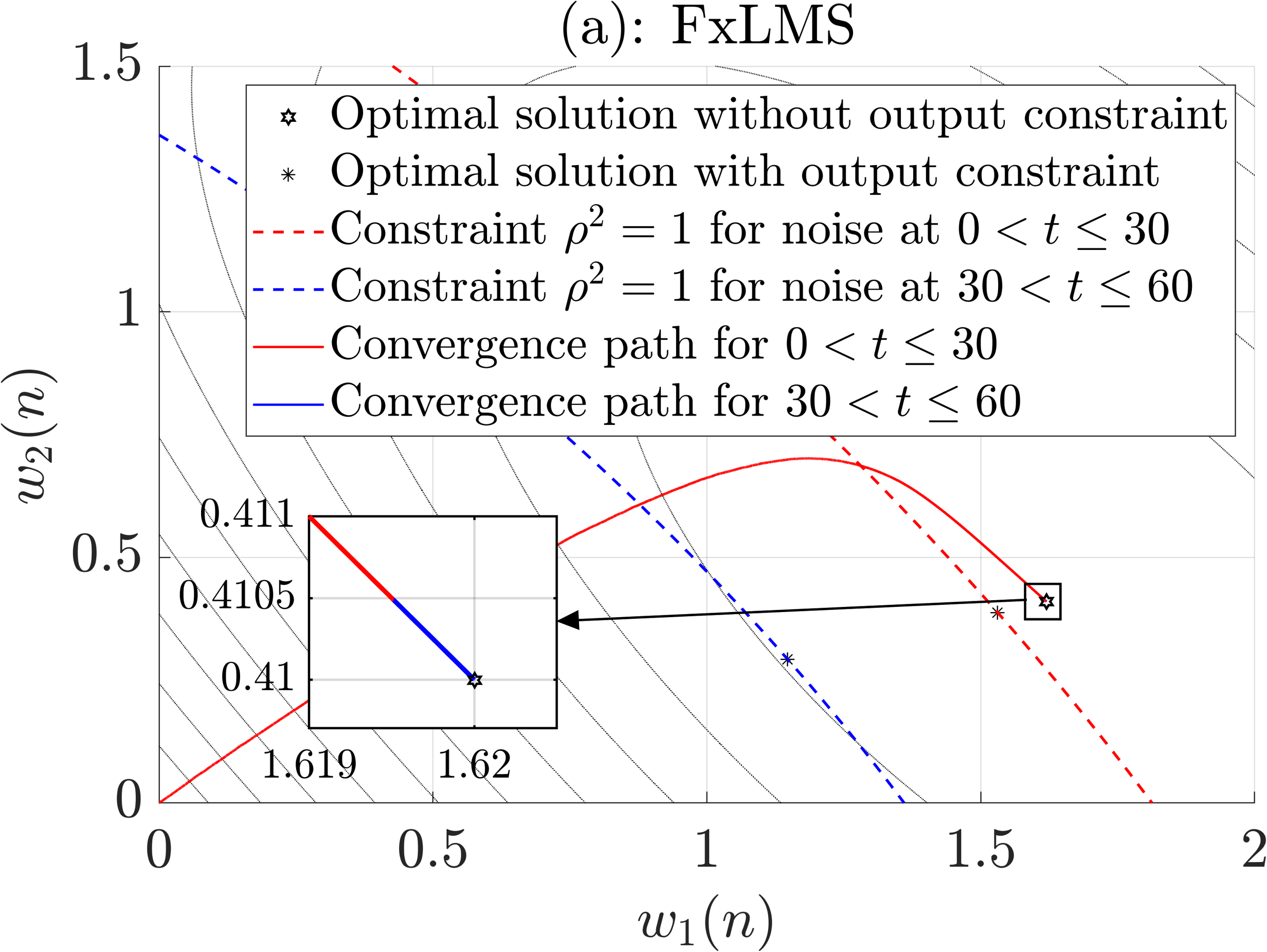} \vspace{-2\baselineskip} 
    \caption{} \label{fig:FxLMS}
    \end{subfigure}
    \begin{subfigure}{0.49\linewidth} \centering
    \includegraphics[width=\linewidth]{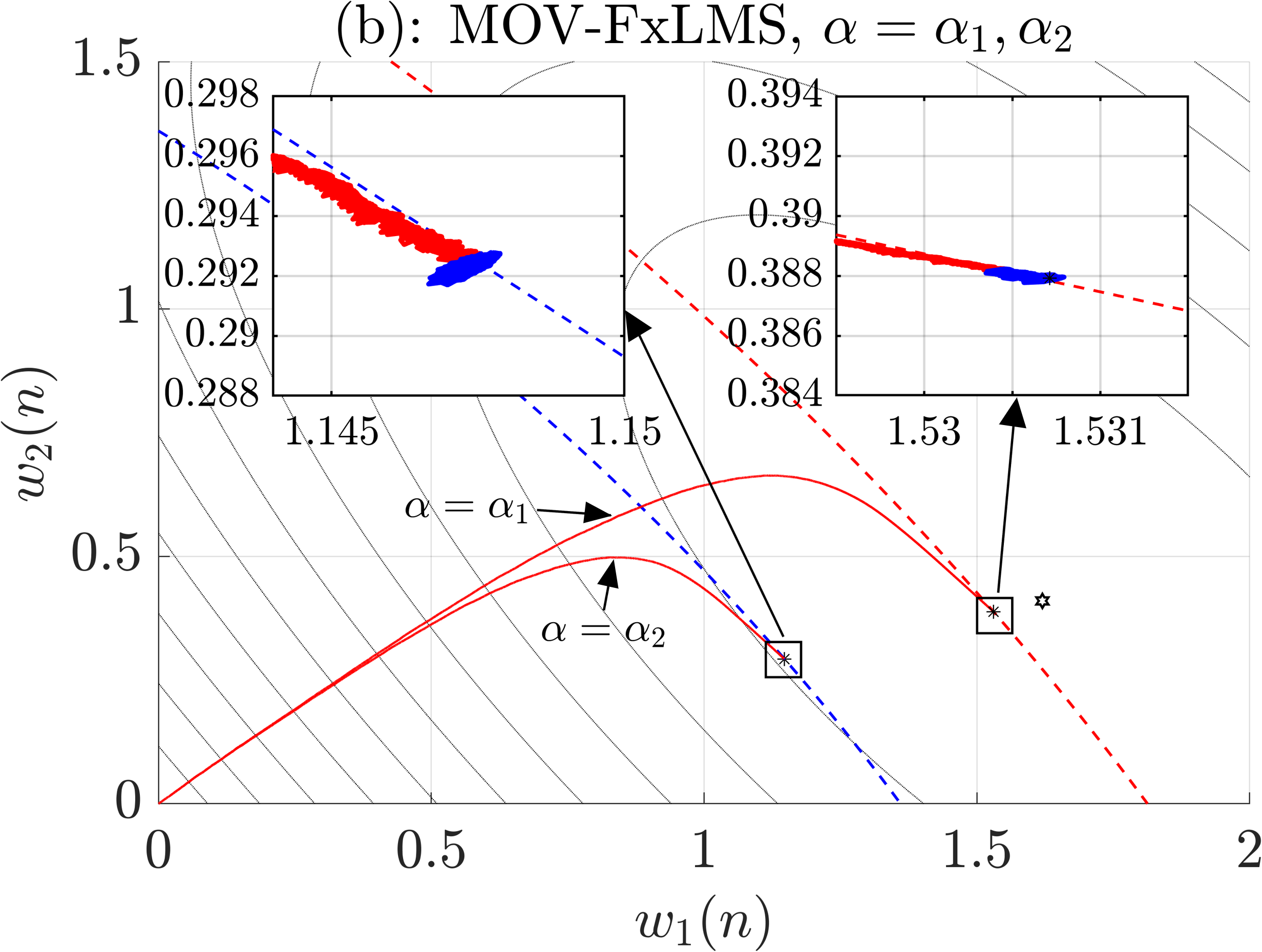} \vspace{-2\baselineskip}
    \caption{} \label{fig:convergence_gamma_both}
    \end{subfigure}
    \begin{subfigure}{0.49\linewidth} \centering
    \includegraphics[width=\linewidth]{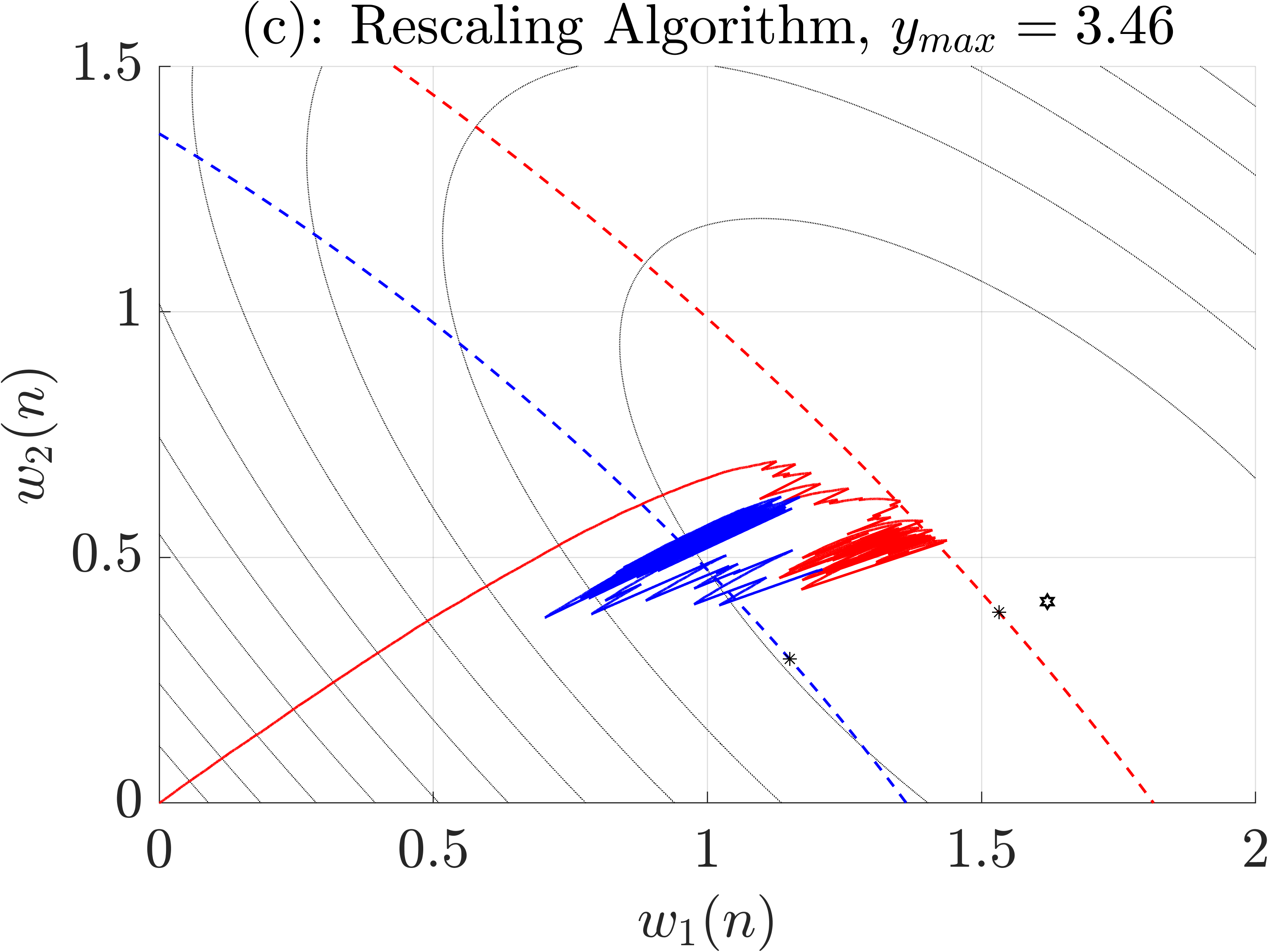} \vspace{-2\baselineskip}
    \caption{} \label{fig:convergence_rescaling}
    \end{subfigure} 
    \begin{subfigure}{0.49\linewidth} \centering
    \includegraphics[width=\linewidth]{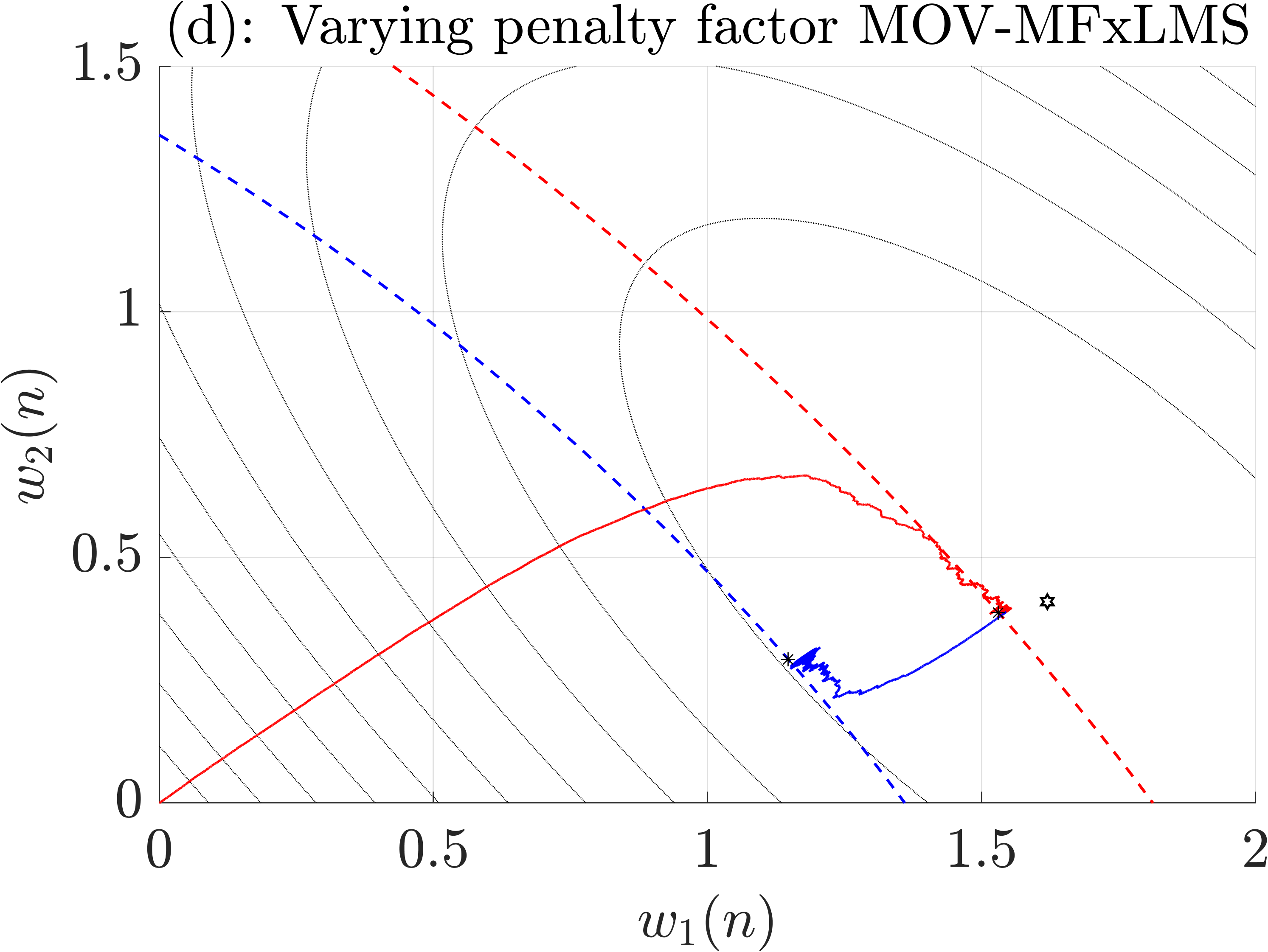} \vspace{-2\baselineskip} 
    \caption{} \label{fig:convergence_vary}
    \end{subfigure}
    \caption{Two-weight convergence paths of \subref{fig:FxLMS}: FxLMS, \subref{fig:convergence_gamma_both}: MOV-FxLMS with optimal penalty factor, \subref{fig:convergence_rescaling}: Rescaling algorithm, and \subref{fig:convergence_vary}: the variable penalty factor MOV-MFxLMS algorithm when dealing with the sudden varying disturbance.} \label{fig:Convergence-path} 
\end{figure}

In this simulation, we compared MOV-MFxLMS with the proposed variable penalty factor, MOV-FxLMS with the predetermined fixed penalty factor, the rescaling algorithm \citep{Qiu2001} with maximum output magnitude parameter of $y_{max} = 3.46$ obtained through trial and error, and the conventional FxLMS algorithm when dealing with a sudden varying disturbance. The reference signal containing a bandlimited white noise (800--\SI{7200}{\hertz}) undergoes a sudden change in its signal power from $\sigma_x^2 = 0.305$ to $\sigma_x^2 = 0.540$ at the \SI{30}{s} mark. 
The optimal penalty factors for the two disturbances are calculated as  $\alpha_1=0.0461$ and $\alpha_2 = 0.3255$, respectively. The sampling rate, step-size, primary and secondary path are set as $F_s = \SI{16}{\kilo\hertz}$, $\mu = 0.0002$, $\mathbf{p} = [1.62 \; 0.41]^\mathrm{T}$ and $\mathbf{s} = [0.03 \; 0.87]^\mathrm{T}$. \fref{fig:Convergence-path} shows the convergence path of the control filter compared between the MOV-MFxLMS, MOV-FxLMS, and the unconstrained FxLMS method. 

As shown in \fref{fig:FxLMS}, FxLMS converges to the optimal control filter $\mathbf{w}_{\text{opt}} = [1.62 \; 0.41]^\mathrm{T}$ while ignoring the subjected power constraint of $\rho^2 = 1$, giving the optimal power output for the control signal of $\sigma_{y_\mathrm{o}}^2 = 1.1214$ for the first \SI{30}{\second} noise and $\sigma_{y_\mathrm{o}}^2 = 1.9769$ for the next \SI{30}{\second}. However, MOV-FxLMS presents an interesting case as only one of the two optimal penalty factors was used throughout the simulation, as shown in \fref{fig:convergence_gamma_both}. It managed to converge to the first optimal constrained filter $\mathbf{w}_{\mathrm{o,1}} = [1.52 \; 0.38]^\mathrm{T}$ in the first \SI{30}{s} when the corresponding $\alpha_1$ was used but failed to converge to the optimal constrained filter in the subsequent \SI{30}{s} with a louder disturbance, violating the power output constraint by having $\sigma_{y_\mathrm{o}}^2 = 1.7630$. On the other hand, the fixed penalty factor of $\alpha_2$ yielded a similar observation where convergence occurred only in the corresponding final \SI{30}{s} where $\mathbf{w}_{\mathrm{o,2}} = [1.14 \; 0.29]^\mathrm{T}$. Although the constraint was not violated in the first \SI{30}{s} with lower amplitude disturbance, it led to an over-constrained output of $\sigma_{y_\mathrm{o}}^2 = 0.5673$. While the rescaling algorithm in \fref{fig:convergence_rescaling} showed a change in its convergence path after the \SI{30}{s} mark, both paths do not converge to their respective optimal constrained filter as it confines the maximum amplitude of the control signal, which is difficult to translate into a precise output-power constraint for a non-periodic noise.
\begin{figure}[t] \centering
    \includegraphics[width=\linewidth]{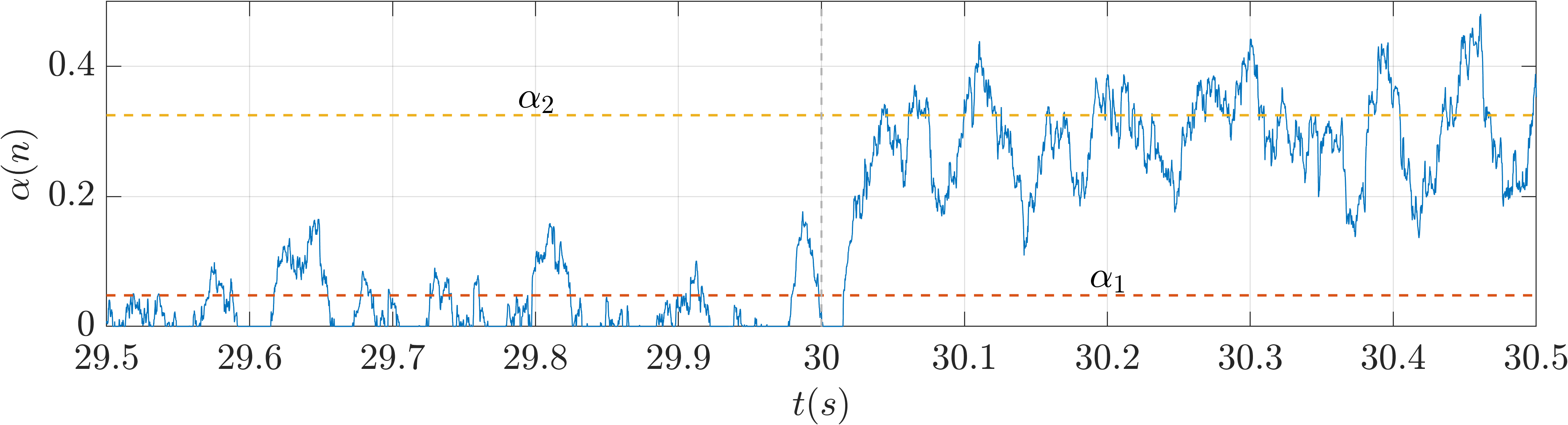}
    \caption{The time history of the variable penalty factor, where the power of disturbance happens to change at $t=\SI{30}{\second}$.} \label{fig:gamma_2coeff} \vspace{-1\baselineskip}
\end{figure}

In contrast, the MOV-MFxLMS algorithm with the variable penalty factor $(K=256)$ can mitigate this problem. As illustrated in \fref{fig:convergence_vary}, it initially converges to the first optimal constrained filter (red path) and shifts to the second optimal constrained filter (blue path) when the second noise is present, thereby satisfying the power-output requirement for the duration of the simulation. The same phenomena can be observed in \fref{fig:gamma_2coeff}, where the calculated penalty factor changes at $t=\SI{30}{\second}$ due to a change in disturbance power. Due to the moving-average operation from \eref{eqn:Gs_est_new}--\eref{eqn:OptimalPenaltyFactorVary}, the variable penalty factor fluctuates around $\alpha_1$ at the first $\SI{30}{\second}$ and around $\alpha_2$ at the next $\SI{30}{\second}$. Therefore, this simulation demonstrates that the proposed variable penalty factor MOV-MFxLMS algorithm is capable to achieve optimal constraint control even if the disturbance has a sudden change in amplitude. The same conclusion can be extended to the variation of the primary noise.   

\subsection{Active control on real noise based on measured paths} 
\label{section:SimulationResultRealMeasurement}
\begin{figure}[!t] \centering
    \begin{subfigure}{\linewidth}    \centering
    \includegraphics[width=0.98\linewidth]{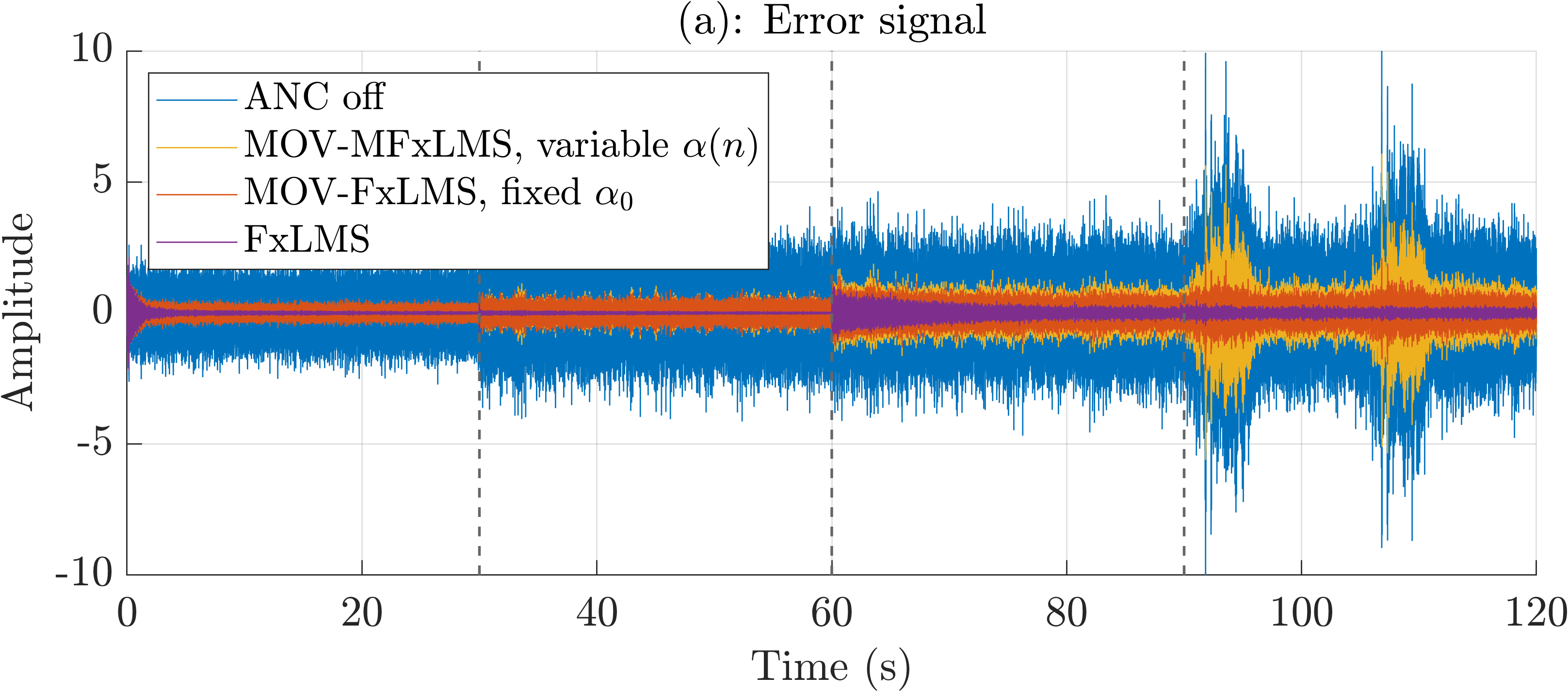} \vspace{-2\baselineskip}
    \caption{} \label{fig:error_changeNoise}
    \end{subfigure} 
    \begin{subfigure}{\linewidth} \centering
    \includegraphics[width=0.98\linewidth]{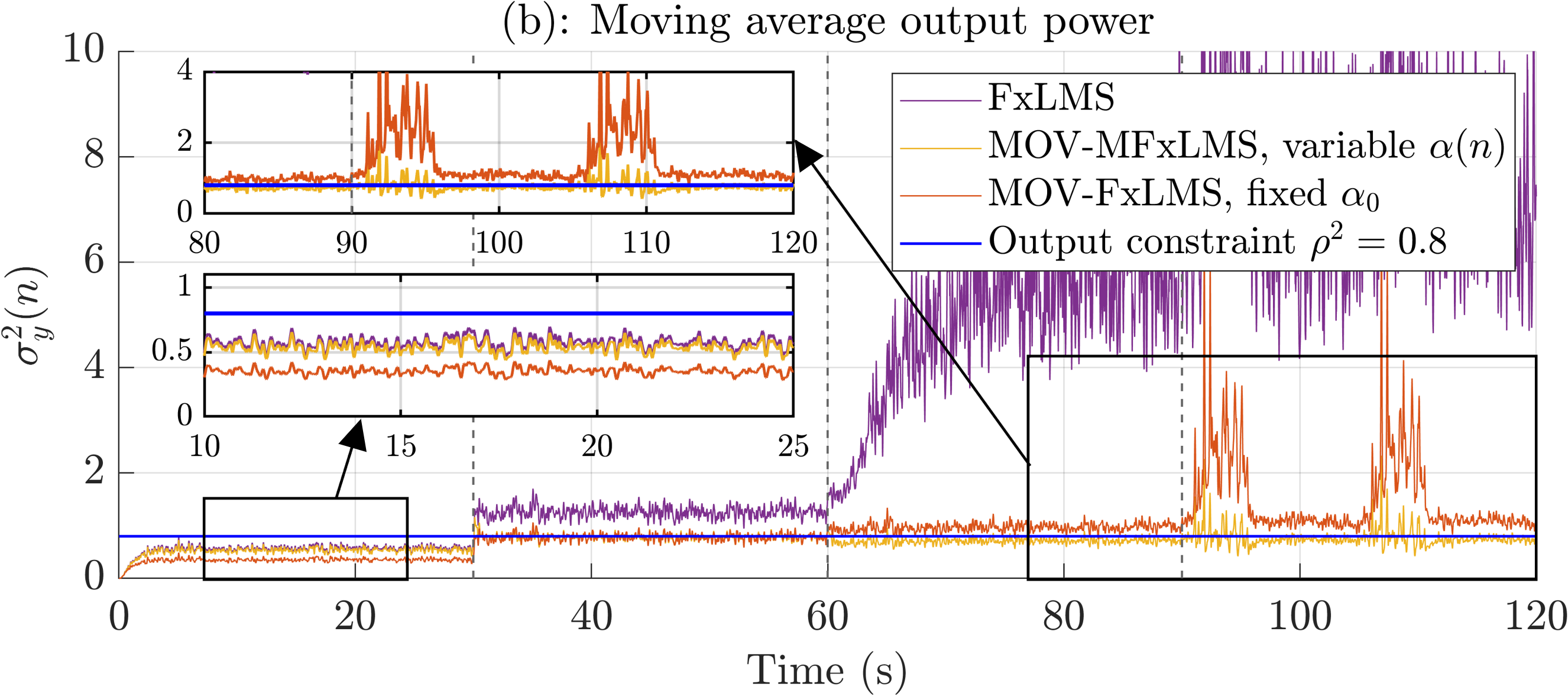} \vspace{-1\baselineskip}
    \caption{} \label{fig:power_changeNoise}
    \end{subfigure}
    \caption{The (a): disturbance and error signal, (b): 1024 samples moving average output power $\sigma_y^2(n)$ during control through various adaptive algorithms with $\rho^2 = 0.8$, $F_s = \SI{16}{\kilo\hertz}$, $I = 256$, $L = 320$, $\mu=0.00005$ and $K = 1024$.} \label{fig:changeRealNoise} \vspace{-1\baselineskip}
\end{figure}

To further verify its effectiveness, the proposed varying MOV-MFxLMS algorithm is utilized to attenuate a compounding noise on the primary and secondary path measured from an air duct \citep{Shi2021b}. \fref{fig:changeRealNoise} depicts the error signal and its moving average output power undergoing four distinct noise stages. The first \SI{30}{\second} contains a 400--\SI{600}{\hertz} broadband noise, and a larger amplitude of 400--\SI{600}{\hertz} broadband noise is compounded to the existing noise for the next \SI{30}{\second}. Similarly, a measured construction noise, having a dominant 70--\SI{400}{\hertz} noise, and aircraft flyover noise, having a dominant 400--\SI{3500}{\hertz} noise, is compounded to its previous noise for the 60--\SI{90}{\second} and 90--\SI{120}{\second} time frame respectively. The output saturation effect in the control signal is not considered in this simulation when exceeding the output power constraint.

The FxLMS algorithm has the most noise reduction as shown in \fref{fig:error_changeNoise} while ignoring its output power constraint shown in \fref{fig:power_changeNoise}, especially in the fourth stage where the output power increases tremendously to an approximate peak value of $\sigma_y^2(n) \approx 10$ due to the non-stationary aircraft noise. For MOV-FxLMS with a fixed penalty factor calibrated for the 30--\SI{60}{\second} noise ($\alpha_0 = 0.1944$), it only managed to meet the power output constraint during the 30--\SI{60}{\second} period but has exceeded the output constraint during the 60--\SI{120}{\second} period due to the additional construction noise and aircraft noise. In contrast, the power output for the 0--\SI{30}{\second} period has been over-constrained as it showed a lower output power than that of the FxLMS algorithm. This issue can be mitigated by the varying penalty factor MOV-MFxLMS algorithm that managed to maintain the power output constraint for the 30--\SI{120}{\second} period as shown in \fref{fig:power_changeNoise}. It has the same convergence behaviour as FxLMS for the first \SI{30}{\second} as the estimated moving averaged disturbance power is less than $K\rho^2\hat{G}_s(n)$, leading to the calculated penalty factor from \eref{eqn:OptimalPenaltyFactorVary} to be 0. This verifies the moving average estimation method from \eqref{eqn:Gs_est_new} and \eqref{eqn:OptimalPenaltyFactorVary}, which satisfy the output power constraint due to a sudden change in noise type.
\vspace{-1\baselineskip}

\section{Conclusion} \label{sec:Conclusion}
In practical ANC applications, a power output constraint is required to prevent undesired output saturation that leads to a degradation in ANC performance. Although the recently proposed MOV-FxLMS algorithm is capable of carrying out this task, it requires prior knowledge of the type of noise and is therefore unsuitable to implement in a real-world scenario. This paper proposes a variable penalty factor that can aid the MOV-MFxLMS algorithm in achieving optimal control under output constraints even with an abruptly changing noise. Numerical simulations based on measured paths and recorded real noise demonstrate the effectiveness of the proposed method in meeting the maximum power output constraint in ANC systems. Since this computational-efficient estimation of this variable penalty factor only requires information on the filtered reference and disturbance, it can be readily realized in real-time ANC applications.


\bibliographystyle{IEEEtranN}
\clearpage
\bibliography{reference} 

\end{document}